\documentclass[11pt]{article}
\usepackage[utf8]{inputenc}
\usepackage[T1]{fontenc}
\usepackage[margin=1in]{geometry}
\usepackage{amsmath,amssymb}
\usepackage{graphicx}
\usepackage{booktabs}
\usepackage{array}
\usepackage[activate={true,nocompatibility},final,expansion=false]{microtype}
\usepackage[hidelinks]{hyperref}
\usepackage{caption}
\title{\textbf{AIx4Soccer: A Unified Platform Architecture for Football Club Management and Structured Athlete Development}}
\author{%
Frederico Falconi Costa$^{1,2,*}$ \quad Salvador Cesar Costa$^{1,2}$ \quad Fabricio F. Costa$^{2,*}$\\[4pt]
\small $^{1}$Empower FC, Belo Horizonte, MG, Brazil\\
\small $^{2}$AIx4Soccer, LLC, Sunnyvale, California, USA\\
\small $^{*}$Corresponding authors: fcosta@aix4all.com, emp.empowerfc@gmail.com
}
\date{\small Preprint: systems/position paper. arXiv category: cs.CY (Computers and Society).}

\newcommand\blfootnote[1]{%
  \begingroup
  \renewcommand\thefootnote{}\footnote{#1}%
  \addtocounter{footnote}{-1}%
  \endgroup
}

\begin{document}
\maketitle
\blfootnote{\textbf{Intellectual-property disclosure.} The platform architecture (AIx4Soccer One Platform), the PDI (Plano de Desenvolvimento Individual) methodology, the Tak Tik marketplace design, and the event-centric semantic data model described in this paper are the subject of a patent application in preparation for filing with the United States Patent and Trademark Office (USPTO). This preprint is published for scientific dissemination and to establish a public, citable disclosure. It does not grant any license to the described intellectual property.}

\begin{abstract}
\noindent Football (soccer) clubs, academies, and federations operate a growing but fragmented portfolio of digital tools: separate systems for video analysis, GPS/performance tracking, medical records, scouting, and administration. This fragmentation is most acute outside the elite European clubs that can afford integration, producing a digital divide that disadvantages grassroots clubs and academies in developing football markets such as Brazil, paradoxically the world's largest exporter of professional players. This paper presents, at a conceptual level, the architecture of ``AIx4Soccer One Platform,'' a multi-tenant cloud SaaS operating system designed to unify club-management workflows and to embed a structured athlete-development methodology, the PDI Framework (``Plano de Desenvolvimento Individual'' / Individual Development Plan). We further describe two companion components: ``Tak Tik,'' a certified two-sided marketplace connecting clubs with certified video analysts under a 75\%/25\% (analyst/platform) revenue split, and the PDI/TBIL methodology, which links individual development plans to video evidence and periodic review cycles. As Materials and Methods, we state the design as explicit requirements and give a formal, implementation-independent specification of the platform's proposed future substrate: an event-centric semantic data model in which every fact is a typed, immutable event in an append-only log that induces a growing knowledge graph. We situate the design against the literature on athlete development frameworks (LTAD, the English FA's EPPP and Four-Corner model, the Belgian and German federation reforms), sports-analytics workflows, two-sided-market economics, and multi-tenant SaaS patterns, and discuss youth data-protection obligations (Brazil's LGPD and the 2025 Digital ECA; the EU GDPR), algorithmic-fairness risks in talent evaluation, and why small, domain-specific models, rather than general-purpose frontier LLMs, are the appropriate intelligence layer. This is a design and early-deployment paper, not an empirical evaluation; we describe what has been built and outline a planned evaluation, making no efficacy claims.\\[4pt]
\end{abstract}

\section{Introduction}\label{sec:intro}
Modern football is a data-intensive enterprise. An elite club's performance department routinely integrates match and training video, event and tracking data, GPS/inertial wearable telemetry, medical and injury records, scouting databases, and athlete-management workflows. Yet these capabilities are delivered by a patchwork of specialized vendors: video analysis and scouting from Hudl and its Wyscout and InStat acquisitions, wearable GPS/load monitoring from Catapult and STATSports, event data from Stats Perform and StatsBomb, and so on. Each is excellent within its niche, but the niches do not compose into a single coherent record of a club or an athlete~\cite{cdf}.

This fragmentation imposes three interacting problems that this paper addresses.

\textbf{Tool fragmentation.} Industry surveys of the football-software landscape consistently describe a market segmented by function (video analysis, scouting, physical monitoring, and administration), with integration left to the buyer. For a resourced Premier League department this is a solvable staffing problem; for the overwhelming majority of the world's clubs it is not. Access itself is stratified: professional scouting platforms such as Wyscout and StatsBomb are described in practitioner guidance as very expensive and typically licensed by clubs rather than individual analysts, pushing independent and lower-budget actors toward cheaper entry-level tools.

\textbf{A professionalization gap.} The management maturity of clubs is highly uneven. Even in Brazil, a football superpower by any sporting measure, the governing body itself has recently moved to ``urgently modernize'' the sport. It restructured competition calendars, launched a financial fair-play model, and professionalized refereeing. In 2025--2026 it also convened a nationwide youth-development working group whose remit explicitly includes the certification and governance of training clubs. The existence of such a working group is itself evidence of a recognized institutional gap between the elite and the base of the pyramid.

\textbf{An athlete-development documentation gap.} The elite European game has converged on structured, documented, multi-dimensional player development. Examples include the English FA's Four-Corner Model (technical/tactical, physical, psychological, social), the Elite Player Performance Plan (EPPP), and the individual development plan (IDP) as an operational instrument. Outside these environments, development is often undocumented, held in a coach's head, and lost when the coach or the player moves. The consequence is not only lost competitive value but lost duty of care. A 2024 University of Essex study led by Dr Jason Moran (published in the \emph{International Journal of Sports Science \& Coaching}) reports that ``just four per cent of talented teen academy prospects make it to the top tier of professional football,'' with only about a further six per cent going on to play in lower leagues~\cite{moran}. The documentation that scaffolds a player's development should therefore also scaffold the education and welfare of the roughly nine in ten who do not turn professional.

These problems are especially consequential in developing football markets. Brazil is the world's largest exporter of players. FIFA's official release on its 2024 Global Transfer Report states that ``Brazilian clubs were in the lead with 1,102 incoming and 1,113 outgoing transfers''~\cite{fifatransfers}, and the CIES Football Observatory Monthly Report n\textsuperscript{o}100 (May 2025) found that Brazil leads all nations with 3,020 expatriate footballers, comfortably ahead of France (2,293) and Argentina (2,171)~\cite{cies}. Yet the Ernst \& Young (EY) study commissioned by the CBF (``Impacto do Futebol Brasileiro,'' 2018 data) counted 360,291 registered athletes across 7,020 clubs, of which only 1,430 were active (874 professional, 556 amateur), spread across 1,499 municipalities, and just 45 clubs held the CBF's ``Certificado de Clube Formador'' (training-club certificate) in one recent year~\cite{ey}. The talent pipeline is vast; the digital and managerial infrastructure beneath it is thin.

\textbf{Contribution.} This paper contributes: (1)~a requirements analysis, grounded in the cited literature and practitioner context, for a unified football club-management platform serving institutional (B2B) users; (2)~a conceptual, non-proprietary architecture for ``AIx4Soccer One Platform,'' a multi-tenant cloud SaaS system with an embedded athlete-development module; (3)~the conceptual design of the ``PDI Framework,'' a structured individual-development methodology mapped to established development-science concepts; (4)~the conceptual design of ``Tak Tik,'' a certified two-sided marketplace for video-analysis labor; and (5)~a formal, implementation-independent specification of an event-centric semantic data model proposed as the platform's future substrate (Sections~\ref{sec:methods} and~\ref{sec:event}). We are explicit that this is a systems/position paper describing a design and an early deployment; it reports no empirical results and makes no efficacy claims. The platform is developed by a Brazil--US partnership: Empower FC and AIx4Soccer, LLC (California, USA).

\section{Related Work}\label{sec:related}
\subsection{The club-management and sports-analytics software landscape}\label{sec:related-tools}
The performance-analysis workflow (capture, tagging/coding, analysis, reporting, and feedback) is well described in the practitioner and applied-science literature~\cite{mackenzie,hughes}, and the video-analyst role has professionalized around it, from live in-match feedback to post-match review and individual player-development clips. The tooling, however, is siloed by function: Hudl for video and coaching workflows (having absorbed Wyscout and InStat for scouting), Catapult and STATSports for wearable GPS and load monitoring, Stats Perform/StatsBomb for event data. Reviews of athlete-monitoring systems note that although data collection is now routine in professional team sport, comparatively little work addresses the end-to-end process of planning, integrating, analyzing, and communicating that data~\cite{thornton}. This is precisely the integration gap a unified platform targets. Recent standardization efforts confirm that significant cross-vendor variation persists in what is collected and how it is represented~\cite{cdf}. Recent web-application-based monitoring studies in youth soccer illustrate both the appetite for lightweight, integrated tools and the engagement challenges they face.

\subsection{Athlete-development frameworks}\label{sec:related-dev}
The Long-Term Athlete Development (LTAD) model, originated by Istvan Balyi and colleagues~\cite{balyi}, organizes development into stages (FUNdamental, Learning to Train, Training to Train, Training to Compete, Training to Win, and lifelong activity) keyed to maturation rather than chronological age. LTAD has been influential and criticized for underweighting contextual factors such as socioeconomic status and coaching quality, and for a physiological emphasis~\cite{ford,bergeron}. In football specifically, national reforms provide the most relevant precedents. The English FA's Four-Corner Model~\cite{fourcorner} frames development across technical/tactical, physical, psychological, and social corners, and underpins the Premier League's EPPP (introduced 2012)~\cite{eppp}, which grades academies into categories 1--4 and has been studied for both its holistic ambitions and its administrative burdens on coaches~\cite{fourcornerstudy,relvas}. Belgium's reform, the ``Golden Vision'' project launched in 2001 after the nation co-hosted Euro 2000, was led by Royal Belgian Football Association technical director Michel Sablon with the consultancy Double PASS~\cite{sablon}; as CNN reported, Sablon's team ``analyzed data from 1,500 youth games, enlisted the cooperation of 70 coaches at all levels of the game and made 120 presentations to the clubs''~\cite{cnn}, standardizing a national playing philosophy, small-sided game formats, and ``parallel teams'' to counter relative age effects. Germany's parallel reform made centrally regulated youth academies a condition of professional licensing for Bundesliga and 2.~Bundesliga clubs from 2002~\cite{dfl}. Underlying these frameworks is a substantial body of skill-acquisition science, notably Ericsson and colleagues deliberate-practice theory~\cite{ericsson1993,ericsson2008}, which prescribes structured, effortful, feedback-rich practice targeted at specific weaknesses, and its subsequent critiques~\cite{macnamara}. The individual development plan (IDP) is the operational unit that translates these frameworks into player-level action, using SMART goals, four-corner assessment, self-reflection for player ownership, and bio-banding by biological maturity~\cite{cumming}. The broader talent-identification and development literature underlines both the multidimensional, dynamic nature of talent and the methodological limitations of one-off selection, reinforcing the case for longitudinal, documented development records~\cite{williams,sarmento,vaeyens}.

\subsection{Two-sided markets and labor-platform economics}\label{sec:related-markets}
The Tak Tik marketplace draws on the economics of two-sided (multi-sided) platforms, formalized by Rochet and Tirole~\cite{rochet2003,rochet2006} and popularized in the platform-strategy literature~\cite{parker}, in which an intermediary must get both sides ``on board'' and where the structure of prices across sides, not only the level, determines participation and volume. Online labor (``gig'') platforms such as Upwork and Fiverr instantiate these dynamics for services, relying on reputation systems, verified skills, and reviews to let strangers assess quality under incomplete information; the HCI and economics literatures document both the value and the biases (including discrimination) of these signals~\cite{whiting}. Certification functions as a stronger, platform-provided quality signal than crowd reputation alone. It is a classic response to quality uncertainty~\cite{akerlof} and operates as a costly signal in the sense of Spence~\cite{spence}.

\subsection{Multi-tenant SaaS architecture}\label{sec:related-saas}
The platform's delivery model rests on well-established multi-tenancy patterns, in which a single application instance serves many tenants with logical data isolation, described along a spectrum from fully isolated (``silo'') to fully shared (``pool'') with hybrid/bridge models in between, and analyzed as encoding business strategy, isolation, and cost trade-offs into software design~\cite{kalra}. This literature provides the vocabulary for describing AIx4Soccer's tenancy at a conceptual level without exposing implementation detail.

\subsection{Positioning}\label{sec:related-positioning}
AIx4Soccer's contribution is not a new analytics algorithm or a new development theory. It is an integration contribution: a single multi-tenant record spanning administration, squad, training, development, and video for institutions that today cannot assemble or afford the elite tool stack, coupled with an explicit methodology (PDI) and a labor marketplace (Tak Tik) that together aim to democratize access to sports-science practice in developing markets. This aligns in spirit with FIFA's Talent Development Scheme, which frames global footballing inequality as the core problem to solve~\cite{fifatds}.

\section{Materials and Methods}\label{sec:methods}
\subsection{Design approach}\label{sec:methods-design}
This work follows a design-science approach~\cite{hevner}: the artifact is the platform design itself, built and refined against explicit requirements derived from the literature of Section~\ref{sec:related} and from practitioner context at the test club, with evaluation planned as staged empirical study (Section~\ref{sec:deployment}). The ``materials'' of this study are therefore (i)~the published literature and public regulatory instruments cited throughout; (ii)~the operational context of a Brazilian club/academy (Empower FC) as design environment; and (iii)~the conceptual artifacts presented here: the requirements (Section~\ref{sec:methods-reqs}), the layered architecture (Section~\ref{sec:arch}), the PDI methodology (Section~\ref{sec:pdi}), the marketplace design (Section~\ref{sec:taktik}), and the formal event model (Section~\ref{sec:methods-formal}). No athlete data were collected or analyzed for this paper.

\subsection{Requirements}\label{sec:methods-reqs}
We synthesize the literature and practitioner context into design requirements. Functional requirements (R) and non-functional requirements (N) are stated so that later sections can be traced back to them.
\begin{itemize}\itemsep2pt
\item \textbf{R1: Single source of truth.} The platform shall maintain one integrated record per club and per athlete spanning administration, squad/roster, training, development, and video, replacing siloed point tools.
\item \textbf{R2: Multi-stakeholder roles.} It shall represent distinct roles (director, coordinator, coach, analyst, athlete, guardian, federation officer) with appropriate, least-privilege permissions.
\item \textbf{R3: Structured athlete development.} It shall operationalize a documented, multi-dimensional development methodology (PDI) with assessment cycles, goal-setting, and review cadence, mapped to established frameworks.
\item \textbf{R4: Video-evidence linkage.} Development goals and assessments shall be linkable to video evidence, so that feedback is concrete and progress is auditable.
\item \textbf{R5: Access to analysis capacity.} The platform shall give clubs that lack in-house analysts access to qualified analysis via a marketplace (Tak Tik), with certification as a quality signal.
\item \textbf{R6: Institutional/federation scale.} It shall support hierarchical, multi-tenant use by clubs, academies, and state/national federations, including oversight and aggregation across subordinate tenants.
\item \textbf{R7: Affordability and low barrier to entry.} The design shall target resource-constrained clubs in developing markets, not only elite buyers.
\item \textbf{R8: Regulatory alignment.} It shall support youth-data-protection obligations (LGPD, Digital ECA, GDPR) and football-governance requirements (e.g., documentation relevant to training-club certification).
\item \textbf{N1: Tenant isolation and security.} Tenants' data shall be logically isolated with least-privilege access control.
\item \textbf{N2: Scalability and availability.} The cloud SaaS deployment shall scale from a single club to federation-wide populations.
\item \textbf{N3: Fairness and transparency.} Any assistive evaluation features shall be designed to mitigate known biases (e.g., relative age effect) and to keep human judgment accountable.
\item \textbf{N4: Interoperability.} The platform shall be able to integrate with, rather than necessarily replace, external data sources where clubs already use them.
\end{itemize}

\subsection{A formal event-centric semantic data model}\label{sec:methods-formal}
We now give an implementation-independent formal specification of the event-centric data model that Section~\ref{sec:event} proposes as the platform's future substrate. The formalism fixes vocabulary and makes the model's properties (immutability, entity linkage through events, unbounded growth, derivability of application state) precise and checkable.

\paragraph{Events.} Let $\mathcal{A}$ denote the set of actors (people, devices, system agents), $\mathcal{V}$ the set of domain entities (athletes, teams, matches, sessions, venues, staff, devices, documents), $\mathcal{T}$ a finite set of event types, $\mathcal{O}$ a set of typed outcomes, and $\mathcal{M}$ a set of media/evidence objects. An \emph{event} is a tuple
\begin{equation}\label{eq:event}
e \;=\; \langle\, \iota,\; \tau,\; \alpha,\; \Sigma,\; t,\; \lambda,\; \omega,\; E \,\rangle,
\end{equation}
where $\iota$ is a globally unique identifier; $\tau \in \mathcal{T}$ is the event type; $\alpha \in \mathcal{A}$ is the actor (\emph{who}); $\Sigma \subseteq \mathcal{V}$ is the non-empty set of subject entities the event is about; $t \in \mathbb{T}$ is the occurrence time (\emph{when}), with $\mathbb{T}$ a totally ordered time domain optionally refined by match/session clocks; $\lambda \in \Lambda$ is the location (\emph{where}), spatial (pitch coordinates) and/or organizational (venue, club, category); $\omega \in \mathcal{O}$ is the outcome; and $E \subseteq \mathcal{M}$ is a (possibly empty) evidence set (video clips, documents, sensor streams) (R4).

\paragraph{Type taxonomy.} $\mathcal{T}$ is organized by a subsumption partial order $\preceq$ (\emph{is-a}) rooted at the abstract type $\mathsf{EVENT}$, with seven top-level families:
\begin{equation}\label{eq:taxonomy}
\tau \;\preceq\; \tau' \;\preceq\; \mathsf{EVENT}, \qquad
\tau' \in \{\mathsf{MATCH}, \mathsf{TRAIN}, \mathsf{MED}, \mathsf{DEV}, \mathsf{VID}, \mathsf{ADM}, \mathsf{MKT}\},
\end{equation}
covering match actions (pass, shot, tackle, goal, \ldots), training events (drill, load measurement, test), medical events (screening, injury, treatment), development events (the PDI stages: assessment, goal-setting, plan, evidence link, review), video events (clip, annotation, tag), administrative events (registration, transfer, contract, consent), and marketplace events (Tak Tik engagement, delivery, settlement). Match subtypes align with established event vocabularies such as SPADL and the Common Data Format~\cite{spadl,cdf}.

\paragraph{The append-only log.} For each tenant $c$ (club, academy, or federation; N1), the system state of record is the \emph{event log}
\begin{equation}\label{eq:log}
L_c(t) \;=\; \big(e_1, e_2, \ldots, e_{n(t)}\big), \qquad t(e_1) \le t(e_2) \le \cdots \le t(e_{n(t)}),
\end{equation}
subject to the \emph{append-only invariant}: for all $t \le t'$,
\begin{equation}\label{eq:appendonly}
L_c(t) \;\sqsubseteq\; L_c(t'),
\end{equation}
where $\sqsubseteq$ denotes the prefix relation. Events are immutable; a correction is itself a new event $e'$ with $\tau(e') = \mathsf{ADM{:}correction}$ and a reference to $\iota(e)$, so the log is simultaneously the audit trail (R8).

\paragraph{The induced knowledge graph.} Entities are related \emph{only through events}. The log induces, at any time $t$, a labeled multigraph
\begin{equation}\label{eq:graph}
G_c(t) \;=\; \big(\mathcal{V}_c,\; R_c(t)\big), \qquad
R_c(t) \;=\; \big\{\, (u, e, v) \;:\; e \in L_c(t),\; u \ne v,\; u, v \in \{\alpha(e)\} \cup \Sigma(e) \,\big\},
\end{equation}
whose edges are events: facts carrying type, time, location, outcome, and evidence. Any derived inter-entity relation is a projection of~\eqref{eq:graph}: for a predicate $\rho$ over events,
\begin{equation}\label{eq:derived}
u \sim_{\rho} v \;\iff\; \exists\, e \in L_c(t) :\; \rho(e) \,\wedge\, (u, e, v) \in R_c(t).
\end{equation}
For example, ``coach $u$ develops athlete $v$'' is the projection with $\rho(e) \equiv \tau(e) \preceq \mathsf{DEV}$; the relation is thereby always dated, attributable, and evidence-backed, and can be recomputed \emph{as of} any past time.

\paragraph{Application state as folds.} In the event-sourcing pattern~\cite{fowler,overeem,kleppmann}, every module view of Section~\ref{sec:arch} (athlete profile, PDI dashboard, squad load view, federation aggregate) is a \emph{projection}: a read model computed by a left fold of a projection function $f_P$ over the log from an initial state $s_0$,
\begin{equation}\label{eq:fold}
S_P(t) \;=\; \mathrm{fold}\big(f_P,\; s_0,\; L_c(t)\big) \;=\; f_P\big(\cdots f_P\big(f_P(s_0, e_1), e_2\big) \cdots, e_{n(t)}\big),
\end{equation}
and temporal (``as-of'') queries follow by folding the truncated log $L_c(t)\vert_{\le t^{*}}$. New views can be defined retroactively over the full history. This property has direct value for compliance reporting and longitudinal development review.

\paragraph{Growth.} The model's asymmetry is the volume argument of Section~\ref{sec:event} made precise: for a club, the entity set is bounded while the log is not,
\begin{equation}\label{eq:growth}
|\mathcal{V}_c(t)| \;=\; O(1) \quad \text{as } t \to \infty, \qquad |L_c(t)| \;=\; \Omega(t),
\end{equation}
with, empirically, a single match contributing on the order of $\mathbb{E}[N_{\mathrm{match}}] \approx 1.6$--$2.0 \times 10^{3}$ recorded events in commercial and open event datasets~\cite{pappalardo}, before counting training, medical, development, video, administrative, and marketplace events.

\paragraph{Event-derived analytics.} Established analytics are functionals of the log, illustrating that event-centric storage subsumes, rather than competes with, current practice. Action valuation in the VAEP framework scores each on-the-ball event by the change in scoring and conceding probabilities it induces~\cite{spadl},
\begin{equation}\label{eq:vaep}
V(e_i) \;=\; \Delta P_{\mathrm{score}}(e_i) \;-\; \Delta P_{\mathrm{concede}}(e_i),
\end{equation}
and recent Large Events Models and soccer foundation models are autoregressive sequence models over event streams~\cite{lem,baron},
\begin{equation}\label{eq:autoregressive}
p_\theta\big(L\big) \;=\; \prod_{i=1}^{n} p_\theta\big(e_i \,\big|\, e_{1}, \ldots, e_{i-1}\big),
\end{equation}
trained, in direct analogy to language models, to predict the next event given the sequence so far. A platform whose native substrate satisfies \eqref{eq:event}--\eqref{eq:growth} is, by construction, the ideal data source for both.

\paragraph{PDI and marketplace instruments.} Within this formalism, an athlete $a$'s PDI is the development-typed sub-log $\mathrm{PDI}_a(t) = \{ e \in L_c(t) : a \in \Sigma(e),\; \tau(e) \preceq \mathsf{DEV} \}$, and a simple cycle-progress metric over review cycle $k$ is
\begin{equation}\label{eq:pdi}
g_a(k) \;=\; \frac{\#\{\text{goals met in cycle } k\}}{\#\{\text{goals reviewed in cycle } k\}} \;\in\; [0, 1],
\end{equation}
computable, with provenance, entirely from events. A Tak Tik settlement event with gross price $p$ splits deterministically as
\begin{equation}\label{eq:split}
\pi_{\mathrm{analyst}} \;=\; 0.75\,p, \qquad \pi_{\mathrm{platform}} \;=\; 0.25\,p,
\end{equation}
the design choice motivated in Section~\ref{sec:taktik}.

\subsection{Evaluation plan}\label{sec:methods-eval}
Consistent with the design-science cycle~\cite{hevner}, evaluation is staged and explicitly deferred: (1)~usability and workflow-fit assessment with staff and athletes at the test club; (2)~adoption and engagement measurement, a known challenge for athlete-monitoring tools (Section~\ref{sec:related-tools}); and (3)~only subsequently, with appropriate study design and ethics review, investigation of development or operational outcomes. Section~\ref{sec:deployment} states the current deployment status; no outcome data exist yet and none are claimed.
\section{Platform Architecture (Conceptual)}\label{sec:arch}
This section describes AIx4Soccer One Platform at a conceptual level only. It intentionally omits proprietary algorithms, code, database schemas, pricing, and client data.

\subsection{Layered architecture}\label{sec:arch-layers}
\begin{figure}[htbp]
\centering
\includegraphics[width=0.95\textwidth]{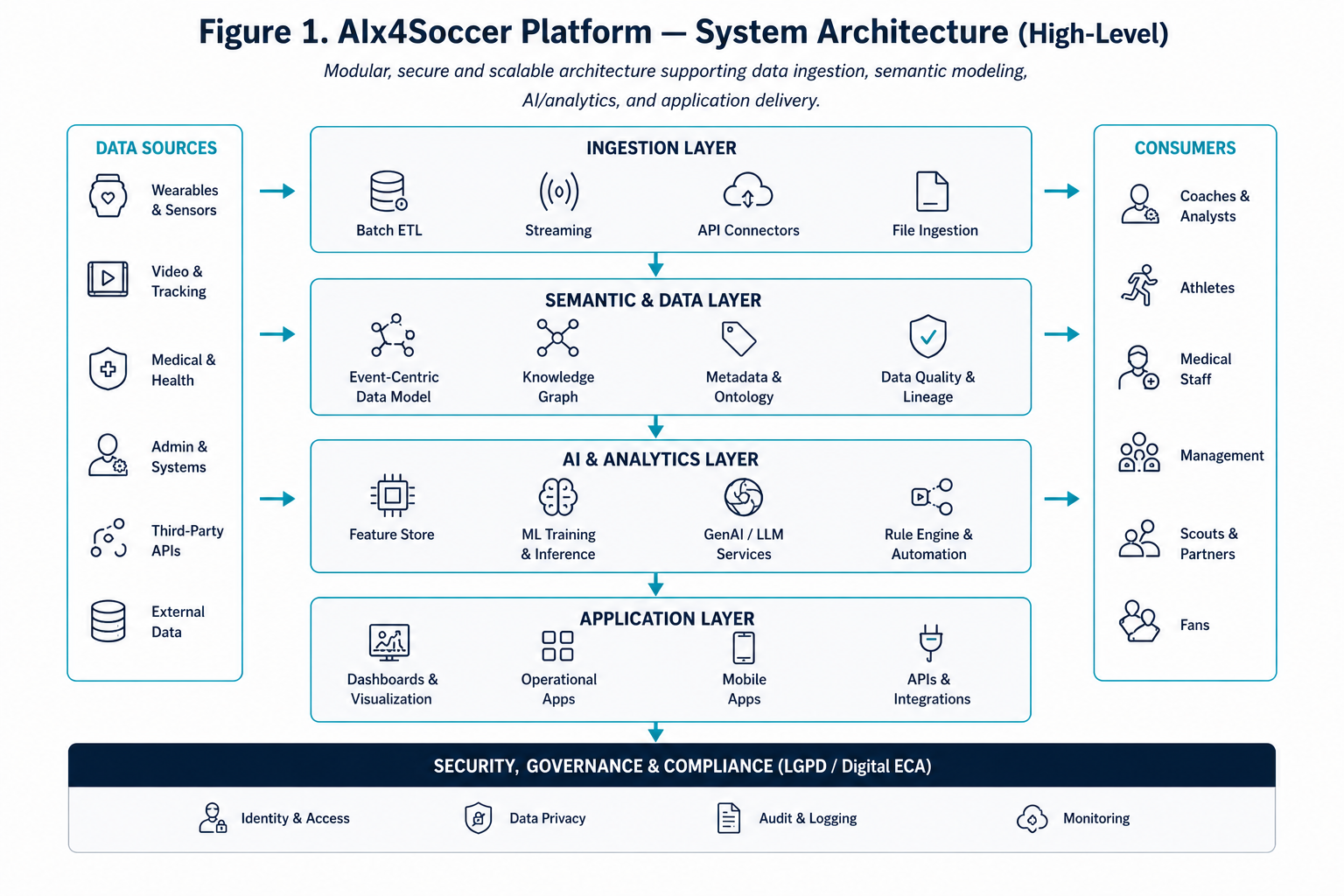}
\caption{Layered architecture of AIx4Soccer One Platform (conceptual). All module reads/writes pass through the tenant-isolated data layer.}
\label{fig:arch}
\end{figure}

The platform follows a conventional layered SaaS structure. The \textbf{presentation layer} provides role-adaptive web and mobile interfaces that render the same underlying record differently for a director, a coach, an analyst, an athlete, or a guardian (R2); federation officers receive aggregate and oversight views (R6). The \textbf{application/module layer} comprises a set of cooperating functional modules (Section~\ref{sec:arch-modules}) implementing club and development workflows (R1, R3). The \textbf{data layer} is a tenant-isolated integrated data store holding the authoritative club and athlete records; all modules read and write against it rather than maintaining private silos (R1, N1). The \textbf{integration layer} provides connectors for selective import/export with external systems clubs may already use (N4) and mediates the marketplace boundary with Tak Tik (R5). The \textbf{platform/infrastructure layer} is cloud-hosted, multi-tenant infrastructure providing isolation, scaling, and availability (N1, N2). Figure~\ref{fig:arch} depicts the stack.

\subsection{Application modules}\label{sec:arch-modules}
Table~\ref{tab:modules} summarizes the application modules.

\begin{table}[htbp]
\centering\small
\caption{Module overview.}
\label{tab:modules}
\begin{tabular}{p{3.4cm}p{5.6cm}p{3.2cm}p{1.6cm}}
\toprule
\textbf{Module} & \textbf{Purpose} & \textbf{Primary roles} & \textbf{Key req.}\\
\midrule
Club Administration & Institutional records, staff, compliance/registration documentation & Director, Coordinator & R1, R8\\
Squad / Roster Management & Teams, categories, registrations, contracts, availability & Coordinator, Coach & R1, R2\\
Training \& Periodization & Session planning, load/training calendars, periodization & Coach, Coordinator & R1, R3\\
Athlete Development / PDI & Individual development plans, assessment cycles, goals, reviews & Coach, Athlete, Guardian & R3, R4\\
Video Analysis Integration & Linking video evidence to athletes/goals; Tak Tik gateway & Analyst, Coach & R4, R5\\
Communication & Structured messaging among stakeholders; review notifications & All roles & R2, R3\\
\bottomrule
\end{tabular}
\end{table}

\subsection{Multi-tenancy and federation hierarchy}\label{sec:arch-tenancy}
Following the tenancy spectrum from the SaaS literature~\cite{kalra}, the platform models each club or academy as a tenant with logically isolated data (N1), formalized as the per-tenant log $L_c$ of Eq.~\eqref{eq:log}. Federations are represented as higher-order tenants able to view and aggregate across their subordinate club tenants for oversight and development-pathway continuity (R6), subject to permissioning. This mirrors the top-down, federation-to-club diffusion of playing philosophy documented in the Belgian and German reforms but implemented as data architecture rather than policy alone.

\subsection{Roles, permissions, and data flow}\label{sec:arch-roles}
\begin{figure}[htbp]
\centering
\includegraphics[width=0.95\textwidth]{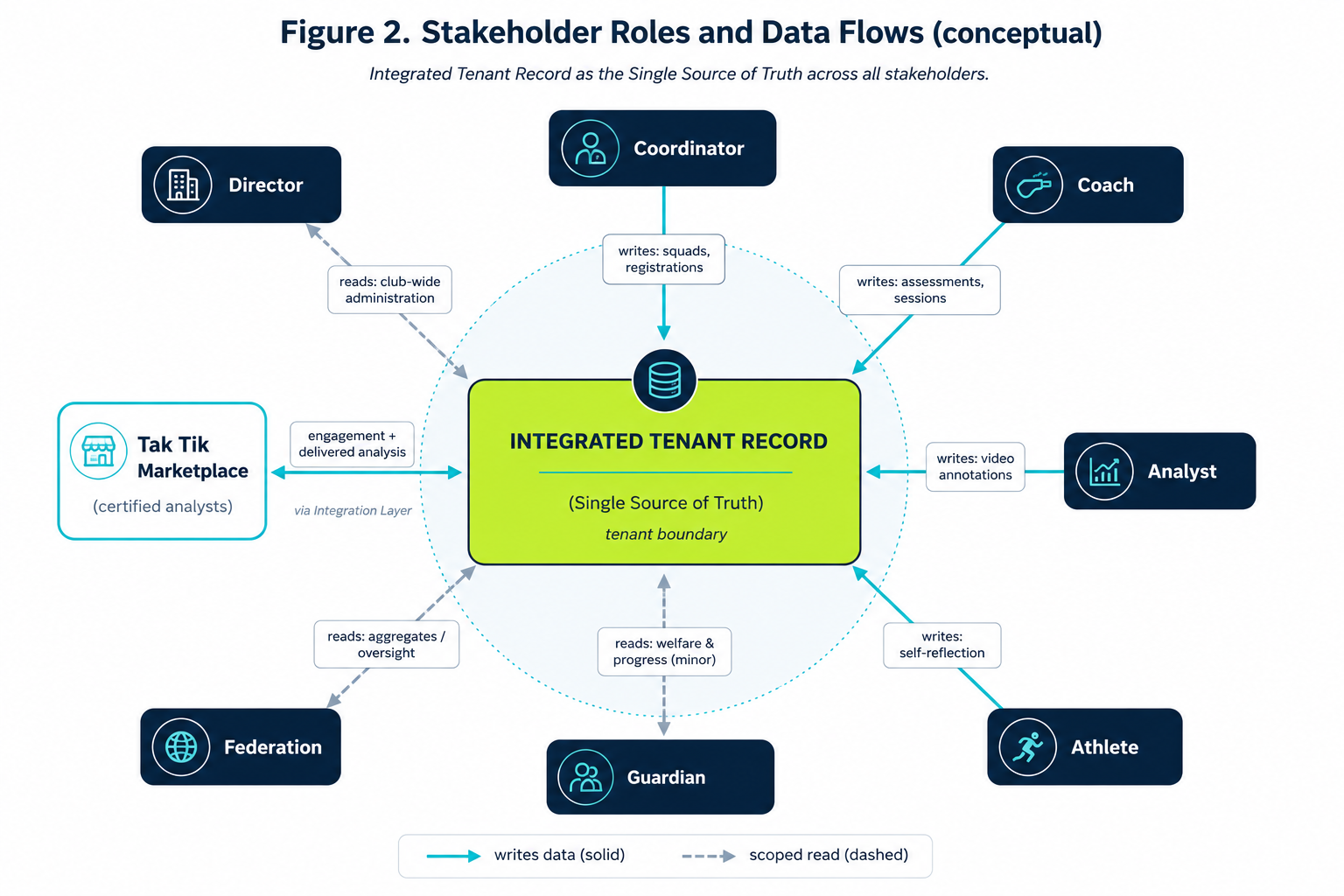}
\caption{Stakeholder roles and data flows (conceptual). Solid arrows denote writes; dashed arrows denote scoped reads. Tak Tik connects through the Integration Layer.}
\label{fig:flows}
\end{figure}

Roles carry least-privilege permissions (R2, N1): guardians see their own athlete's development and welfare information; coaches see their squads; analysts see the material they are engaged to analyze; directors and coordinators see club-wide administration; federation officers see aggregate/oversight data. The athlete is a first-class participant in their own development record (consistent with IDP ``player ownership''). Figure~\ref{fig:flows} shows the stakeholder data flows around the integrated record.

\subsection{Comparison with the fragmented status quo}\label{sec:arch-compare}
Table~\ref{tab:compare} contrasts today's fragmented tool categories with the unified approach. The point is not that AIx4Soccer replaces specialized analytics vendors, but that it provides the connective record those tools currently lack, and that it brings development documentation, today the least-tooled category, into first-class support.

\begin{table}[htbp]
\centering\small
\caption{Fragmented tool categories vs.\ unified platform. Vendor examples are drawn from the cited market literature and illustrate categories, not endorsements.}
\label{tab:compare}
\begin{tabular}{p{3.1cm}p{3.6cm}p{3.2cm}p{4.0cm}}
\toprule
\textbf{Capability category} & \textbf{Typical dedicated tools} & \textbf{Integration burden today} & \textbf{AIx4Soccer approach}\\
\midrule
Video analysis / coding & Hudl, Nacsport, Dartfish & Separate logins, exports & Native module + marketplace (R4, R5)\\
Scouting / match database & Wyscout, StatsBomb, InStat & Costly, club-licensed & Out of scope as data vendor; integrable (N4)\\
Physical / GPS load & Catapult, STATSports & Separate hardware/software & Integrable via connectors (N4)\\
Medical / availability & Point solutions & Manual reconciliation & Part of integrated record (R1)\\
Athlete development / IDP & Ad hoc documents, spreadsheets & Uncaptured, coach-held & PDI module (R3, R4)\\
Administration & Generic office tools & Disconnected from sporting data & Club administration module (R1)\\
\bottomrule
\end{tabular}
\end{table}

\section{The PDI Framework}\label{sec:pdi}
The \textbf{PDI Framework} (``Plano de Desenvolvimento Individual'' / Individual Development Plan), also referred to as the \textbf{PDI/TBIL methodology}, is the platform's structured athlete-development methodology. It is presented here as a conceptual methodology; no proprietary content is disclosed.

\subsection{Rationale and grounding}\label{sec:pdi-grounding}
PDI operationalizes the IDP concept documented in the sport-pedagogy literature and used by elite academies: a per-athlete plan combining multi-dimensional assessment, specific measurable goals, player self-reflection, and periodic review. It is grounded in three literatures: (a)~multi-dimensional development frameworks (the FA Four-Corner Model's technical/tactical, physical, psychological, social dimensions~\cite{fourcorner}; LTAD's maturation-keyed staging~\cite{balyi}); (b)~deliberate-practice theory (specific, effortful, feedback-rich work on identified weaknesses~\cite{ericsson1993}); and (c)~the video-evidence and monitoring practices of applied performance analysis~\cite{thornton,bourdon,mackenzie}.

\subsection{The PDI cycle}\label{sec:pdi-cycle}
\begin{figure}[htbp]
\centering
\includegraphics[width=0.72\textwidth]{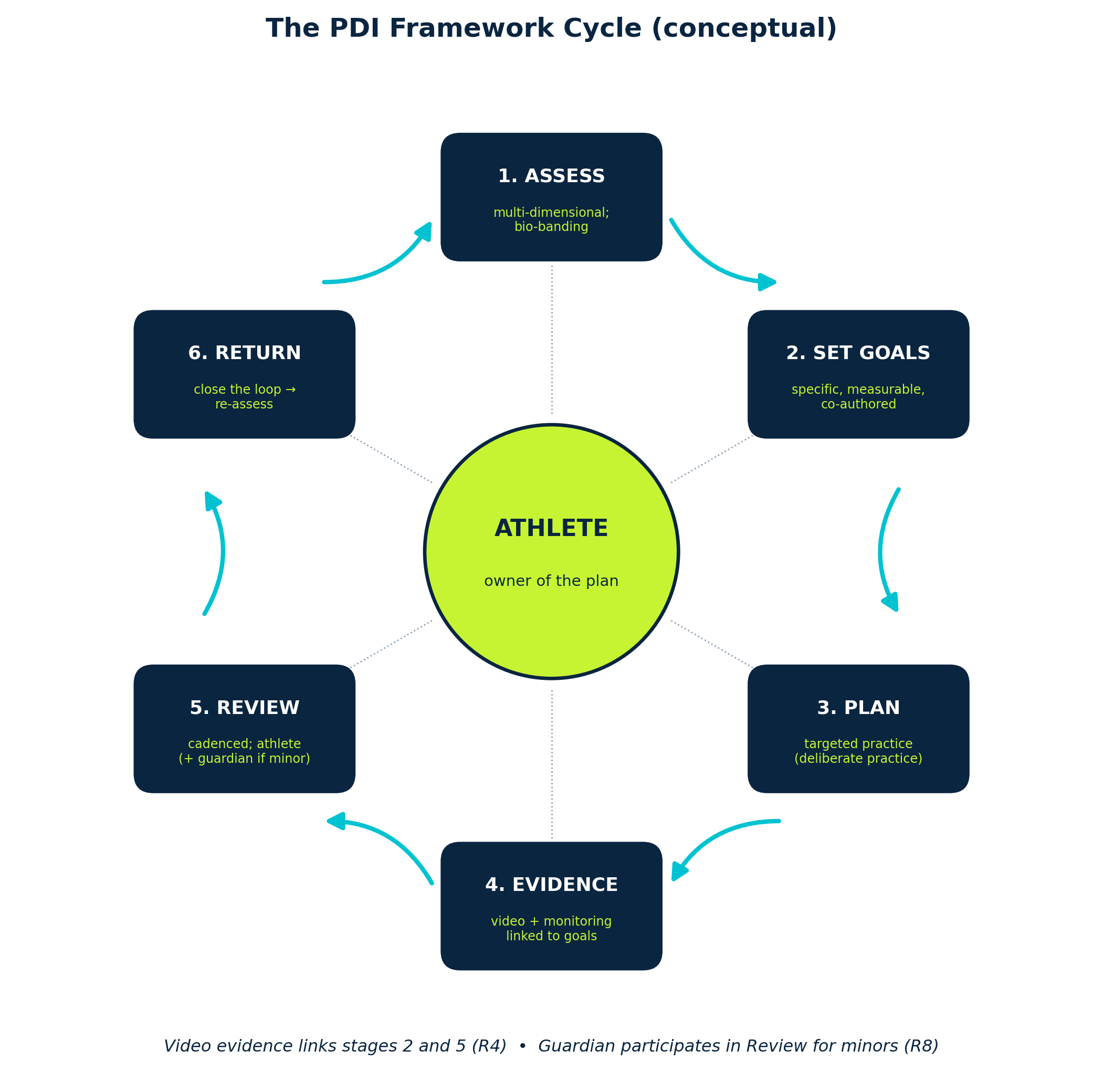}
\caption{The PDI cycle (conceptual). The athlete sits at the hub as owner of the plan; video evidence links goal-setting and review; guardians participate in reviews for minors.}
\label{fig:pdi}
\end{figure}

PDI is a recurring cycle rather than a one-off document (R3):
\begin{enumerate}\itemsep2pt
\item \textbf{Assess} the athlete across development dimensions, using biological as well as chronological age where relevant (bio-banding~\cite{cumming}) to reduce maturation bias.
\item \textbf{Set goals} that are specific and measurable, co-authored with the athlete to build ownership.
\item \textbf{Plan} targeted training/practice addressing prioritized goals.
\item \textbf{Evidence} progress by linking video and monitoring data to specific goals (R4).
\item \textbf{Review} on a defined cadence with the athlete (and guardian, for minors), updating goals.
\item \textbf{Return} to assessment, closing the loop.
\end{enumerate}
In the formalism of Section~\ref{sec:methods-formal}, the cycle is the development-typed sub-log $\mathrm{PDI}_a$, with cycle progress measurable as in Eq.~\eqref{eq:pdi}. Figure~\ref{fig:pdi} depicts the cycle.

\subsection{Mapping PDI to established concepts}\label{sec:pdi-mapping}
Table~\ref{tab:pdimap} maps the PDI stages to the established development-science concepts of Section~\ref{sec:related-dev}. The ``TBIL'' designation denotes the methodology's internal pedagogical structuring; its detailed content is proprietary and is deliberately not disclosed here. The claim of this paper is only that the shape of PDI (cyclical, multi-dimensional, video-evidenced, athlete-owned) is consistent with, and traceable to, the cited development science.

\begin{table}[htbp]
\centering\small
\caption{PDI stages mapped to established development-science concepts.}
\label{tab:pdimap}
\begin{tabular}{p{6.2cm}p{7.6cm}}
\toprule
\textbf{PDI stage} & \textbf{Established concept / source literature}\\
\midrule
Assess (multi-dimensional; bio-banding) & FA Four-Corner Model~\cite{fourcorner}; LTAD maturation staging~\cite{balyi}; relative-age-effect research~\cite{cobley}\\
Set Goals (specific, measurable, co-authored) & SMART goal-setting in IDP practice; deliberate-practice specificity~\cite{ericsson1993}\\
Plan (targeted practice) & Deliberate practice (Ericsson et al.)~\cite{ericsson1993}\\
Evidence (video + monitoring linkage) & Performance-analysis feedback workflows~\cite{mackenzie}; athlete-monitoring systems~\cite{thornton}\\
Review (cadenced, athlete/guardian) & IDP review and player-ownership practice; EPPP holistic review~\cite{eppp}\\
Continuity across pathway & Federation development-pathway models (Belgium, Germany)~\cite{cnn,dfl}\\
\bottomrule
\end{tabular}
\end{table}

\section{Tak Tik: A Certified Video-Analysis Marketplace}\label{sec:taktik}
\textbf{Tak Tik} is a two-sided marketplace connecting clubs (demand side) with certified video analysts (supply side).

\subsection{The gap it addresses}
Section~\ref{sec:related} established that professional analysis capacity is unevenly distributed: elite clubs employ full-time analysts, while smaller clubs and academies, especially in developing markets, often cannot. Simultaneously, there is a pool of trained and aspiring analysts (a professionalizing labor market) without efficient access to paying institutional work. Tak Tik is designed to clear this two-sided matching problem (R5).

\subsection{Two-sided-market design}
Following Rochet--Tirole~\cite{rochet2003,rochet2006}, a marketplace must bring both sides on board and must attend to the structure of value distribution across sides, not merely its level. Tak Tik's publicly stated design choice is the \textbf{75\%/25\% revenue split} of Eq.~\eqref{eq:split}: 75\% to the analyst and 25\% to the platform. The split is deliberately skewed toward the supply side to attract and retain qualified analysts. This in turn makes the platform valuable to the demand side (the cross-side network effect). This split is a design parameter disclosed as a matter of positioning; no other pricing is disclosed.

\subsection{Certification as a quality signal}
\begin{figure}[htbp]
\centering
\includegraphics[width=0.92\textwidth]{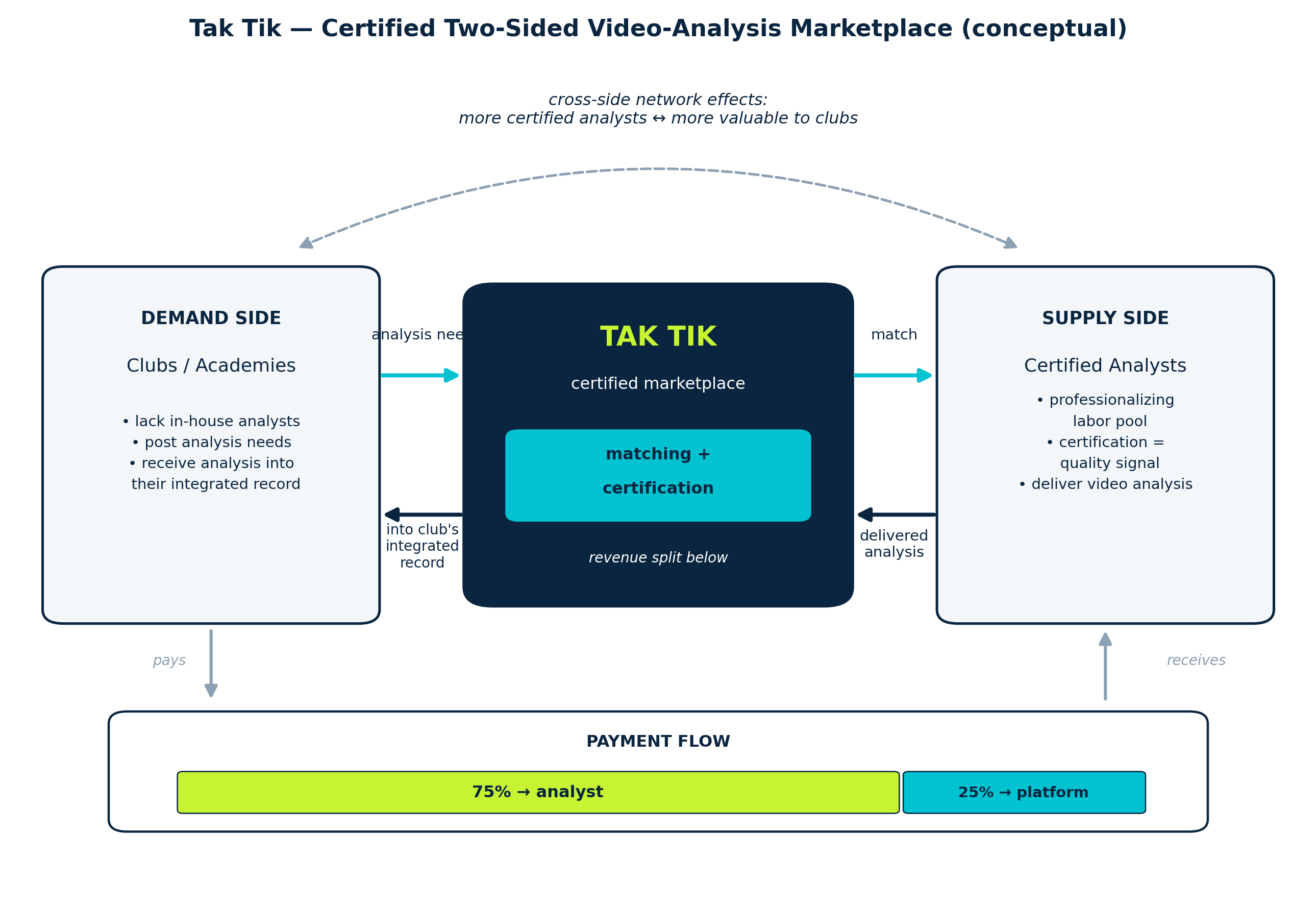}
\caption{Tak Tik two-sided market (conceptual): matching and certification at the center, cross-side network effects, and the 75\%/25\% analyst/platform payment split.}
\label{fig:taktik}
\end{figure}

General gig platforms rely on crowd reputation and reviews, signals that are useful but subject to bias~\cite{whiting}. Tak Tik instead uses \textbf{certification} as a platform-provided quality signal layered on top of reputation, mitigating the adverse-selection problem that arises under quality uncertainty~\cite{spence,akerlof}. Certification addresses the incomplete-information problem (clubs cannot easily assess a stranger analyst's competence) more strongly than ratings alone, and it aligns with the broader standardization gap in the analyst labor market noted in Section~\ref{sec:related-tools}. Reputation/review mechanisms can complement certification but are explicitly secondary to it. Figure~\ref{fig:taktik} depicts the marketplace.

\subsection{Coupling with the platform}
Because Tak Tik delivers analysis directly into the club's integrated record and can link outputs to PDI goals (R4), it is not a standalone freelancing site but a capability extension of the platform: the marketplace supplies labor, and the platform supplies the context (the athlete record, the development plan) that makes the labor actionable.

\subsection{The whole model at a glance}\label{sec:wheel}
\begin{figure}[htbp]
\centering
\includegraphics[width=0.8\textwidth]{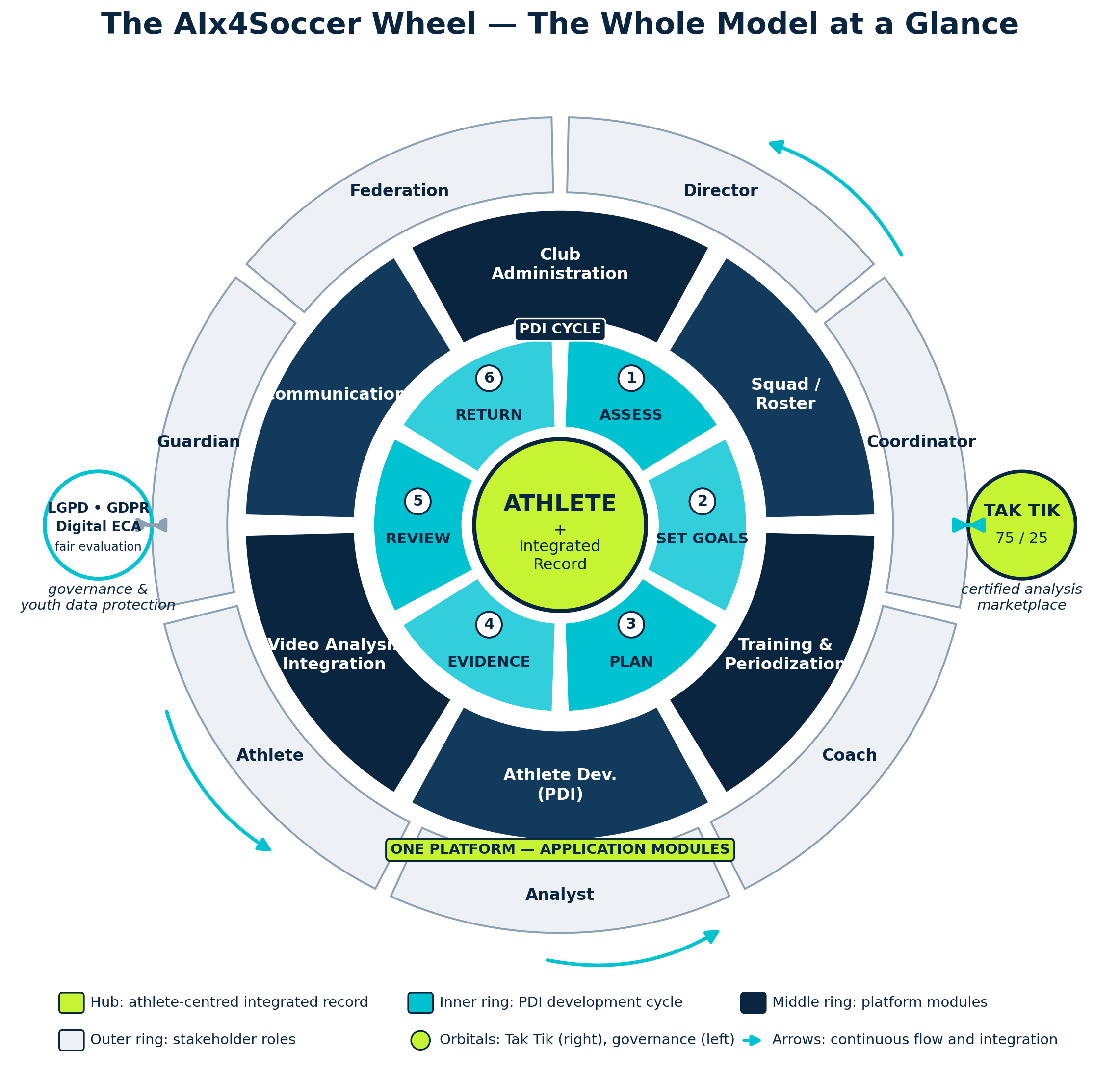}
\caption{The AIx4Soccer Wheel: athlete-centred hub, PDI cycle (inner ring), platform modules (middle ring), stakeholder roles (outer ring), with Tak Tik and governance as orbitals.}
\label{fig:wheel}
\end{figure}

Figure~\ref{fig:wheel} summarizes the complete model as a wheel. At the hub sits the athlete and their integrated record; the inner ring is the PDI development cycle that continuously turns around the athlete; the middle ring comprises the six platform modules that operationalize club workflows; and the outer ring holds the stakeholder roles who read from and write to the record under least-privilege permissions. Two orbitals complete the picture: Tak Tik (right), the certified analysis marketplace that feeds labor into the wheel, and the governance layer (left), comprising LGPD, Digital ECA, GDPR, and fair-evaluation constraints, which bounds how the wheel may turn.

\section{Towards an Event-Centric Semantic Data Model}\label{sec:event}
The architecture of Section~\ref{sec:arch} is deliberately conventional: modules over a shared record. In this section we propose what we regard as the platform's future data model: the semantic topology formalized in Section~\ref{sec:methods-formal}, in which the central concept is not the athlete, the match, or the club, but the \textbf{EVENT} of Eq.~\eqref{eq:event}. This is a proposal and a design direction, not a description of the current implementation.

\subsection{Why events are the natural center}\label{sec:event-why}
Football is, at bottom, a stream of events. The sports-analytics community already models the match this way: commercial and open event-data specifications record every pass, shot, duel, and foul with position, time, actor, and outcome~\cite{pappalardo,cdf}, and representation languages such as SPADL normalize heterogeneous vendor catalogs into a uniform on-the-ball action vocabulary precisely so that events become analyzable units~\cite{spadl}. Open tooling (e.g., the kloppy library) and the recently proposed Common Data Format exist because the field converged on the event as the lingua franca of match data, while suffering from vendor fragmentation around it~\cite{cdf,kloppy}. Ontological work in the same direction, such as RDF-based football ontologies and knowledge graphs populated from match event feeds, demonstrates that events map naturally onto graph semantics~\cite{cdfonto}.

The volume argument is decisive, and Eq.~\eqref{eq:growth} states it precisely. A single match yields on the order of 1{,}600--2{,}000 recorded events; the open Wyscout corpus alone contains roughly three million events for one season of seven competitions~\cite{pappalardo}, and research corpora used for action-valuation models span more than ten million actions across a few seasons~\cite{spadl}. A club's daily life multiplies this further: every training drill, load measurement, medical screening, PDI assessment, video annotation, registration, and marketplace delivery is also an event. Whereas entity tables saturate (a club has only so many athletes), the event stream grows without bound. It is the platform's naturally accumulating asset.

\subsection{The proposed topology}\label{sec:event-topology}
\begin{figure}[htbp]
\centering
\includegraphics[width=\textwidth]{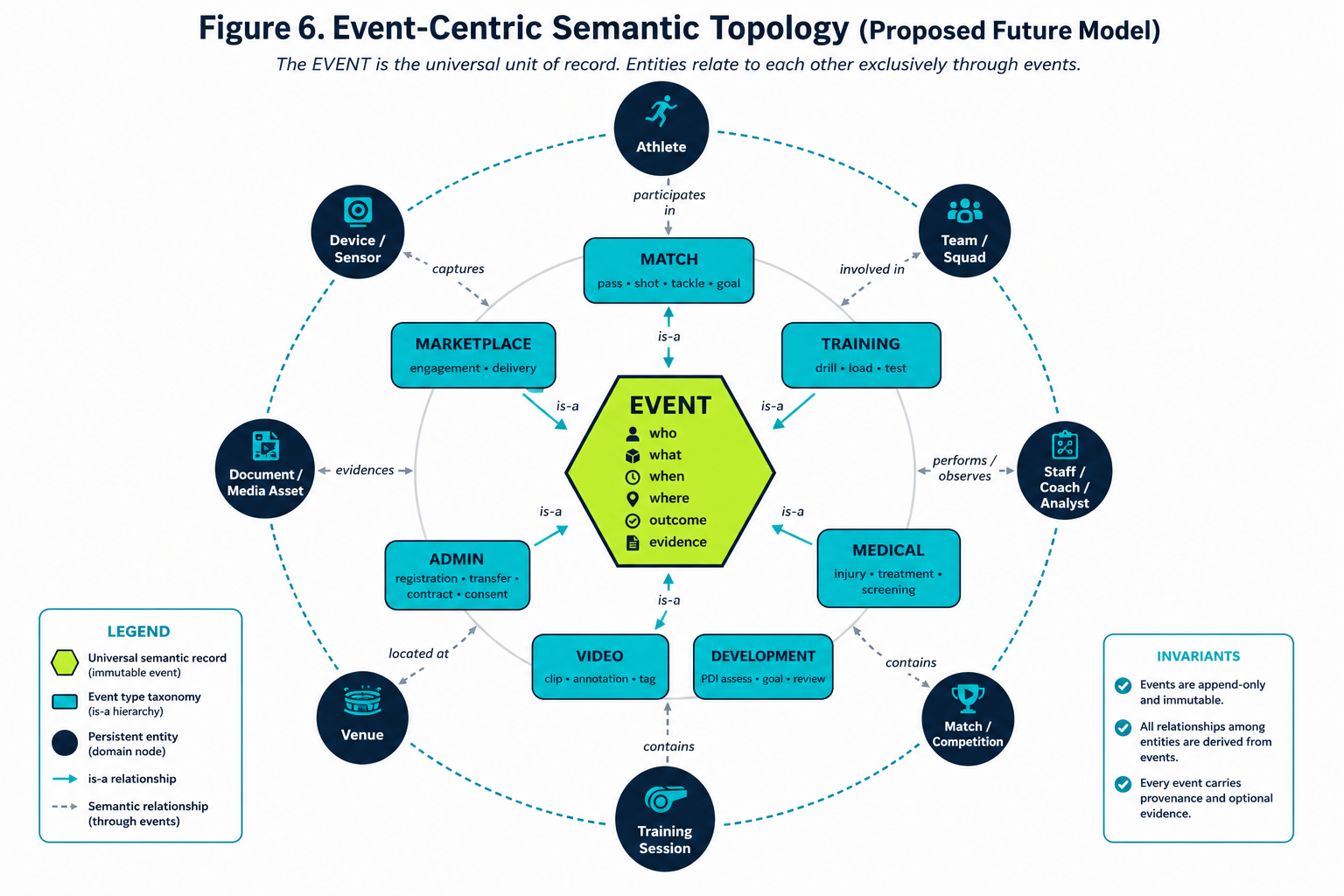}
\caption{Proposed event-centric semantic data model: the EVENT (hexagon) as the universal unit of record with who/what/when/where/outcome/evidence semantics; an event-type taxonomy (inner ring); entities linked to one another only through events (outer ring); an append-only log that grows into a knowledge graph as the platform is populated. Formal counterpart: Eqs.~\eqref{eq:event}--\eqref{eq:growth}.}
\label{fig:eventmodel}
\end{figure}

Figure~\ref{fig:eventmodel} sketches the proposed semantic topology, whose formal counterpart is Eqs.~\eqref{eq:event}--\eqref{eq:derived}. At the center sits the EVENT as the universal unit of record, carrying five semantic slots: who (actor), what (typed action), when (timestamp), where (spatial or organizational location), and evidence (an optional link to video, document, or sensor data), plus an outcome. Around it, an \textbf{event-type taxonomy} (Eq.~\eqref{eq:taxonomy}) specializes the concept: match events, training events, medical events, development events (PDI assessments, goal-settings, reviews), video events, administrative events (registrations, transfers, contracts), and marketplace events (Tak Tik engagements and deliveries). Entities (athlete, team, match, session, venue, staff, device) are nodes of the graph that connect to one another only \emph{through} events (Eq.~\eqref{eq:graph}): an athlete and a venue are related because events occurred there; a coach and a player are related through the assessments, sessions, and reviews that link them (Eq.~\eqref{eq:derived}). The result is a knowledge graph whose edges are facts with provenance and time, in the spirit of graph-based integration efforts in football and of spatio-temporal sports-data research generally~\cite{cdfonto,gudmundsson}.

\subsection{Architectural analogue: the append-only event log}\label{sec:event-log}
Software engineering offers a mature architectural analogue: \textbf{event sourcing}, in which system state is not stored as mutable current values but derived by replaying an append-only log of immutable domain events~\cite{fowler}. This is exactly the fold semantics of Eq.~\eqref{eq:fold} under the invariant of Eq.~\eqref{eq:appendonly}. Empirical studies of industrial event-sourced systems document the pattern's benefits (complete auditability, temporal ``as-of'' queries, and retroactive read models built from history) alongside its real costs in schema evolution~\cite{overeem}, and the append-only log is a foundational abstraction of modern data-intensive systems generally~\cite{kleppmann}. These properties align remarkably well with the platform's obligations: an immutable event history is simultaneously a development record (what was assessed, decided, and reviewed, and when), a compliance audit trail (who accessed or changed what, relevant to LGPD/GDPR accountability), and a defensible evidence base for training-compensation and formation-certification regimes that reward documented contribution over time (Section~\ref{sec:disc-gov}).

\subsection{What a populated event graph enables}\label{sec:event-enables}
Finally, the event-centric model is forward-compatible with the most interesting current direction in soccer analytics: treating the match, and potentially the whole life of a club, as a \emph{language of events}. Action-valuation frameworks such as VAEP already derive player contribution values purely from event streams (Eq.~\eqref{eq:vaep})~\cite{spadl}. More recently, Large Events Models and transformer-based foundation models for soccer are trained, in direct analogy to large language models, to predict the next event given the sequence so far (Eq.~\eqref{eq:autoregressive}), and are proposed as general simulation and analytics backbones~\cite{lem,baron}. A platform whose native substrate is a well-typed, ever-growing event log is, by construction, the ideal data source for such models. In the long run this includes models that learn not only match dynamics but development dynamics: which sequences of training, assessment, and review events precede successful athlete outcomes. Interoperability principles (FAIR: findable, accessible, interoperable, reusable) apply directly, arguing for typed, well-described events with stable identifiers so that the graph remains machine-actionable as it grows~\cite{fair}. We propose this event-centric semantic model as the platform's north star for data architecture, to be adopted incrementally as the current module-over-record design is populated.

\section{Deployment Context and Current Status}\label{sec:deployment}
This is an early-deployment, honest-status section; it reports no efficacy metrics and invents no results.

AIx4Soccer is developed by a Brazil--US partnership. \textbf{Empower FC} serves as the Brazilian operational partner, the initial real-world environment in which the club-administration, squad, training, and PDI workflows are exercised. \textbf{AIx4Soccer, LLC} (California, USA) is the technology partner. The intended user base is institutional and B2B: football clubs, academies, and state and national federations, not individual consumers.

At the time of writing, the work described is a platform design with an early deployment in a single Brazilian club/academy context. We make no claims about adoption scale, retention, competitive outcomes, or development efficacy. The staged evaluation plan of Section~\ref{sec:methods-eval} applies; we flag explicitly that no outcome data exist yet and that this paper should not be read as evidence of effectiveness.

\section{Discussion}\label{sec:disc}
\subsection{Democratizing sports science in developing markets}\label{sec:disc-demo}
The platform's guiding thesis is that the binding constraint for most of the football world is not analytical sophistication but integration and access. Brazil exemplifies the paradox: it is the largest exporter of players in the world~\cite{fifatransfers,cies}, yet its base is under-tooled and under-professionalized, and its own governing body is actively working to certify and modernize training clubs. A unified, affordable, federation-scalable record, coupled with a methodology (PDI) and a labor marketplace (Tak Tik), is a plausible lever precisely because it lowers the barrier that specialized elite tools raise. This aligns with FIFA's framing of global footballing inequality as the problem to solve~\cite{fifatds}, though we claim alignment of intent, not demonstrated impact.

\subsection{Youth data protection}\label{sec:disc-privacy}
Because the primary subjects are minors, data protection is central, not peripheral (R8). Brazil's \textbf{LGPD} (Lei Geral de Proteção de Dados, Lei n\textsuperscript{o}~13.709/2018) requires that processing of children's and adolescents' data be conducted in their \textbf{best interest}, and, for children under 12, with \textbf{specific and prominent consent} from at least one parent or legal guardian, with controllers making reasonable efforts to verify that consent (Art.~14)~\cite{lgpd}. Brazil's \textbf{Digital Statute of the Child and Adolescent} (``Digital ECA,'' Law No.~15.211/2025), in force from March 2026, adds design-stage obligations and prohibits profiling for targeted advertising to minors~\cite{eca}. This is directly relevant to any platform holding youth-athlete data. The EU \textbf{GDPR} provides a comparable but not identical regime~\cite{gdpr}; commentators note the LGPD tends to be more consent-protective for children, while the GDPR permits legitimate-interest bases with balancing tests. LGPD's extraterritorial reach means a US technology partner processing Brazilian minors' data is squarely within scope. Design implications include guardian-mediated consent flows (represented as first-class consent events in the model of Section~\ref{sec:methods-formal}), least-privilege access (N1), purpose limitation, data-minimization, and records of processing. We present these as obligations the architecture must support, and note that full compliance is an ongoing operational responsibility, not a solved problem.

\subsection{Football-governance alignment}\label{sec:disc-gov}
The platform's documentation orientation is congruent with Brazilian formation regulation. The \textbf{Lei Pelé} (Lei n\textsuperscript{o}~9.615/1998), as amended by Lei n\textsuperscript{o}~12.395/2011, provides at Art.~29 that the training entity may sign a player's first professional contract from age 16 for a term of up to five years~\cite{leipele}. The CBF's \textbf{Certificado de Clube Formador}, regulated in January 2012 (following the 2011 amendment) and updated by later resolutions, attests which clubs meet the legal requirements for the proper technical and social formation of athletes in Brazil, and entitles a certified club to pursue the domestic training indemnity (``indenização por formação'')~\cite{ccf}. Internationally, FIFA's Regulations on the Status and Transfer of Players provide that training compensation is paid to a player's training club(s) when a player signs a first professional contract and on transfers up to the end of the calendar year of the player's 23rd birthday (Art.~20 and Annexe~4), while up to 5\% of a transfer fee is distributed as a solidarity contribution to clubs that trained the player between ages 12 and 23 (Art.~21 and Annexe~5)~\cite{rstp}. Structured, auditable development records of the kind PDI produces, immutable and datable under the append-only invariant of Eq.~\eqref{eq:appendonly}, are exactly the evidence base such regimes reward, though we are careful to note that the FIFA solidarity mechanism attaches to the player's sporting passport rather than to the CCF, and we do not overstate the platform's role in any legal entitlement.

\subsection{Algorithmic fairness in talent evaluation}\label{sec:disc-fair}
Any system that assists talent evaluation risks encoding bias. The \textbf{relative age effect} is the over-selection of players born early in the selection year, documented across large youth-soccer samples and confirmed meta-analytically~\cite{cobley}. It is the canonical example, and the Belgian reform's ``parallel teams''~\cite{cnn} and bio-banding~\cite{cumming} were explicit countermeasures (N3). In a German talent-program dataset of 16{,}138 U12--U15 players, both subjective and objective assessments of performance were shown to be biased by relative age, with a skewed distribution toward early-born players~\cite{romann}. Our design stance is that assistive features should surface maturation and birth-quarter context, avoid opaque single-score talent rankings, and keep human judgment accountable; we do not claim to have solved algorithmic fairness, and we treat it as a first-order design constraint and an open research problem.

\subsection{Why general-purpose frontier models are not a substitute and why small, domain-specific models fit}\label{sec:disc-slm}
A natural objection to this entire enterprise is that frontier AI laboratories (e.g., OpenAI, Anthropic, Google DeepMind) could simply replicate it: their large language models (LLMs) exhibit broad, near-human competence across tasks~\cite{bommasani}, and ``football knowledge'' is abundantly represented in their web-scale training corpora. We take this objection seriously and argue that it conflates two different kinds of knowledge and two different kinds of product.

\textbf{The knowledge that matters here is not in the training distribution.} What frontier models absorb from the web is football's \emph{public} discourse: rules, history, punditry, and published analytics. What this platform manages is football's \emph{private} operational record: which drill athlete $A$ performed on Tuesday, what load her sensor registered, what her coach assessed in the last PDI review, which video clip evidences her progress against goal three. These facts are proprietary, longitudinal, and club-specific. Crucially, they do not exist anywhere until a system of record like this one captures them. No amount of pre-training scale can recover data that was never written down or never shared; the moat is the populated event graph of Section~\ref{sec:event}, not model capability. A general model can be granted access to such data via retrieval or fine-tuning, but that presupposes exactly the structured, well-typed, consented data substrate whose construction is this paper's subject: the platform is upstream of any model, frontier or otherwise.

\textbf{System-of-record duties are hostile to generative behavior.} LLMs are prone to hallucination: fluent, plausible, unfounded output. This failure mode is extensively documented and only partially mitigable~\cite{ji}. In a system whose subjects are predominantly minors and whose records feed development decisions, welfare duties, and (Section~\ref{sec:disc-gov}) formal training-compensation evidence, fabricated content is not an inconvenience but a safeguarding and legal hazard. The platform's core must therefore be deterministic and auditable (the append-only event log of Section~\ref{sec:event-log}), with any generative assistance strictly grounded in, and traceable to, recorded events. Data-protection law compounds this: LGPD's best-interest standard and the Digital ECA's design-stage obligations for minors' data (Section~\ref{sec:disc-privacy}) weigh heavily against routing children's medical, biometric, and developmental records through third-party model APIs whose retention and training practices the club does not control.

\textbf{Small language models (SLMs) are the architecturally correct fit.} The evidence is now substantial that, for narrow and repetitive task distributions, small models fine-tuned on domain data match or exceed frontier generalists: carefully curated training data lets sub-4B-parameter models rival vastly larger ones~\cite{phi1,phi3}; large-scale empirical studies report fine-tuned small models outperforming GPT-4 on most tested specialized tasks~\cite{loraland,bucher}; and a prominent position paper argues that SLMs, being sufficiently powerful, operationally more suitable, and drastically more economical, are the natural engine of agentic systems that perform ``a small number of specialized tasks repetitively and with little variation''~\cite{belcak}, which is a precise description of club workflows (summarize this session, draft this PDI review from these events, tag this clip). SLMs can be deployed within the tenant boundary, on-premises or in the platform's own cloud. This keeps minors' data inside the LGPD compliance perimeter, at inference costs one to two orders of magnitude below frontier APIs, and with an alignment surface small enough to audit. Notably, the most relevant models in football analytics already follow this pattern: the Large Events Models and the soccer foundation model of Section~\ref{sec:event-enables} are compact, domain-specific transformers trained on event streams, not frontier LLMs~\cite{lem,baron}.

Our position is therefore not that frontier laboratories lack the capability to build such systems, but that the binding constraints favor a vertical platform. These constraints are proprietary longitudinal data, institutional relationships with clubs and federations, regulatory posture toward minors' data, certification of a human labor market, and unit economics at grassroots price points. They favor a platform whose intelligence layer is built from small, domain-specialized models trained on its own event graph. In the heterogeneous-system framing of~\cite{belcak}, frontier models may still play a peripheral role for open-ended language tasks; the system of record, and the models that learn football's operational language from it, remain domain-native.

\subsection{Limitations}\label{sec:disc-limits}
This paper has clear limitations. It is a design and position paper: the architecture is described conceptually and without proprietary detail, so it cannot be independently reproduced from this text. There is no empirical evaluation and therefore no evidence of usability, adoption, or efficacy. The deployment is early and confined to a single Brazilian club context, limiting external validity. Claims about the value of integration and of the PDI/Tak Tik design are argued from the literature and practitioner reasoning, not demonstrated. Finally, the founders' involvement means this account is not disinterested; we have tried to constrain claims accordingly.

\section{Conclusion and Future Work}\label{sec:conclusion}
We have presented a conceptual architecture for AIx4Soccer One Platform: a multi-tenant cloud SaaS operating system for football club management with an embedded, literature-grounded athlete-development methodology (the PDI Framework) and a certified two-sided marketplace for video-analysis labor (Tak Tik, with a 75\%/25\% analyst/platform split). We accompany it with a formal specification of the event-centric semantic data model proposed as its future substrate. The design responds to three documented problems (tool fragmentation, a professionalization gap, and an athlete-development documentation gap) that fall hardest on clubs and academies outside the elite European game, and it is aimed squarely at institutional users in developing football markets such as Brazil.

The contribution is integrative and methodological rather than algorithmic, and the claims are correspondingly modest: this is what has been designed and begun to be built, grounded in the relevant science, and it is not yet evidence that the approach works. Future work is therefore primarily empirical: staged usability and adoption studies at the test club; measurement of engagement over time; careful, ethics-reviewed study of whether structured, video-evidenced development documentation changes practice; and evaluation of the marketplace's ability to clear the analyst matching problem and of its certification as a quality signal. Additional work should address fairness auditing of any assistive evaluation features, interoperability with incumbent data vendors, and the operational realities of LGPD/Digital ECA/GDPR compliance for youth-athlete data at federation scale. Architecturally, the priority future direction is the event-centric semantic data model of Sections~\ref{sec:methods-formal} and~\ref{sec:event}: incrementally migrating the platform's substrate to a typed, append-only event graph that unifies match, training, medical, development, video, administrative, and marketplace events. As it is populated, the graph becomes both the club's institutional memory and the training ground for the next generation of event-sequence models in football. Consistent with Section~\ref{sec:disc-slm}, we expect that intelligence layer to be built from small, domain-specialized models trained within the platform's compliance perimeter, rather than delegated to general-purpose frontier APIs.

\paragraph{Author's note on scope and honesty.} This is a systems/position paper. The AIx4Soccer platform, the PDI/TBIL methodology, and the Tak Tik marketplace are described only at a conceptual level. No proprietary algorithms, source code, database schemas, internal financial data, pricing (other than the publicly stated 75\%/25\% marketplace split), or client identities are disclosed. All quantitative claims about the football industry, existing tools, development frameworks, regulation, and player-development statistics are attributed to the cited public sources; no empirical results are claimed for the platform itself. The platform architecture, the PDI methodology, the Tak Tik marketplace design, and the event-centric semantic data model are the subject of a patent application in preparation for filing with the United States Patent and Trademark Office (USPTO); this preprint constitutes a public scientific disclosure and grants no license to that intellectual property.

\end{document}